\documentclass[aps,twocolumn,showpacs,superscriptaddress]{revtex4}

\newcommand{\BC}{k_{\rm B}}

\usepackage{graphicx}
\usepackage{bm}
\begin{document}
\title{Role of conformational entropy in force-induced bio-polymer unfolding}
\author{Sanjay Kumar} 
\affiliation{Department of Physics, Banaras Hindu University,
Varanasi 221 005, India }
\author{Iwan Jensen}
\affiliation{ARC Centre of Excellence for Mathematics and Statistics of Complex Systems,
Department of Mathematics and Statistics, 
The University of Melbourne, Victoria 3010, Australia}
\author{Jesper L. Jacobsen}
\affiliation{Universit\'e Paris Sud, UMR8626, LPTMS, F-91405 Orsay Cedex, France \\
   Service de Physique Th\'eorique, CEA Saclay, F-91191 Gif-sur-Yvette, France
} 
\author{Anthony J. Guttmann}
\affiliation{ARC Centre of Excellence for Mathematics and Statistics of Complex Systems,
Department of Mathematics and Statistics, 
The University of Melbourne, Victoria 3010, Australia}


\begin{abstract}
A statistical mechanical description of flexible and semi-flexible
polymer chains in a poor solvent is developed in the constant force and 
constant distance ensembles. We predict the existence of many 
intermediate states at low temperatures stabilized by the force. 
A unified response to pulling and compressing forces has been obtained
in the constant distance ensemble. We show  the signature of a cross-over
length which increases linearly with the chain length. 
Below this cross-over length, the
critical force of unfolding decreases with temperature, while above,
it increases with temperature. For stiff chains, we report for the
first time  ``saw-tooth" like behavior in the force-extension curves which has
been seen earlier in the case of protein unfolding.
\end{abstract}
\maketitle

During the past decade force has been used as a thermodynamic variable 
to understand molecular interactions and their role in the structure of 
bio-molecules \cite{rief,busta}. By exerting a force in the pN range
one can experimentally study the elastic, mechanical, 
structural and functional properties of bio-molecules \cite{busta1}. The 
dependence of force on concentration, pH of the solvent, loading rate
and temperature provides 
basic understanding of the interactions \cite{evan,bloom,evans}. 
Many biological reactions involve large conformational
changes which provide well defined mechanical reaction co-ordinates,
{\it e.g.} the end-to-end distance of a polymer, that can be used 
to follow the progress of the reaction \cite{busta1}. Such 
processes have been modeled by a simple two state model \cite{busta1}.
The applied force ``tilts" the free energy surface along the
reaction co-ordinate by an amount linearly dependent on
the  end-to-end distance. The kind of transitions induced by the
applied force  are the folding-unfolding transition of proteins \cite{rief}, 
the stretching and unzipping transition of dsDNA \cite{bhat,bbs} or  
the ball-string transition of a polymer \cite{haupt}. From polymer theory
we know that a polymer chain will, depending on temperature, be in either a 
collapsed state or a swollen state \cite{degennes}.  The  end-to-end distance 
$\langle R \rangle$ scales as $N^\nu$, where $N$ is the chain length and $\nu$ is 
the end-to-end distance exponent. In the collapsed state  (low temperatures) 
$\nu=1/d$, while at high temperatures $\nu$ is given quite accurately by 
the Flory approximation $\nu=3/(d+2)$ \cite{degennes} (actually this formula 
is exact for $d=1$, 2 and 4). 
It should be noted that by varying temperature 
alone, a polymer can not acquire the conformation of a stretched state 
where $\nu=1$. Hence force not only ``tilts" the free energy surface but also induces
a new ``stretched state" which is otherwise not accessible. 
Moreover, recent experiments suggest that there are many 
intermediate states involved which
are crucial to the understanding  of unfolding experiments 
and lie beyond the scope of two state models \cite{haupt,
hanbin,lemak}.

The non-equilibrium thermodynamics of small systems has mostly
been studied in the ``constant force ensemble" ({\bf CFE}) 
where the control parameter is the average extension.  
Most applications of atomic force microscopy apply the force
using a linear ramp protocol with a very small velocity.
Such systems may be considered 
to be in quasi-static equilibrium and the appropriate ensemble is
 the ``constant distance ensemble" ({\bf CDE}).
In the thermodynamic limit both ensembles are expected to give similar 
results \cite{busta2}. However, single molecule experiments study systems of 
finite size and the results may  depend on the ensemble \cite{zema}.  
Apart from this, the physical 
constraints imposed by experimental setups have not been fully taken 
into account. For example, in atomic force microscopes, receptor 
and ligand molecules are attached to a substrate and a transducer, 
respectively.  The loss of  entropy due to the confinement  
has been ignored in most models.  

Theoretically these transitions have been studied using simple models 
such as freely jointed chain (FJC) or worm like chain (WLC) models \cite{fixman,doi}. 
The WLC model has been
used to study the force-extension curves of bio-molecules.
However, this model ignores excluded volume effects and attractive interactions
between chain segments and is thus only well suited for 
modeling the stretching of polymers in a good solvent \cite{doi}. 
For a polymer in a poor solvent, the force-extension 
curve shows the existence of a plateau region at
a well defined force \cite{haupt,hanbin}. Experimental observations
of polymers in poor solvents and an improved theoretical understanding
of semi-flexible polymers have given rise to theories of globules
with well defined internal structure \cite{grass,garel}. These
theories have potential applications in the study and understanding of 
the basic mechanisms of protein folding. 

The purpose of this communication is firstly to provide a complete
phase diagram using exact results of finite chains. However, 
we focus our studies on the behavior of the force-extension curves 
at low temperatures where the usual Monte Carlo approach and the
theoretical models discussed above fail. These studies  have direct
relevance to phenomena seen in experiments with single biomolecules. 
We report the existence of intermediate states stabilized by 
the force at low temperatures. We also show the
signature of a crossover length, below which the transition temperature 
decreases with force, while above the crossover length the transition 
temperature increases with force. When bending rigidity is taken into 
account, we predict ``saw-tooth'' like oscillations in the regime of 
large polymer stiffness, as observed in protein unfolding experiments.

\begin{figure}[t]
\includegraphics[width=3.2in]{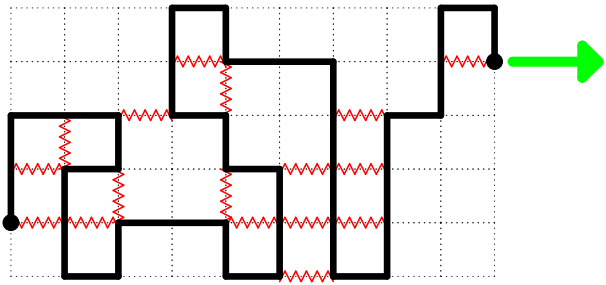}
\caption{\label{fig:model} An ISAW on the square lattice with one end
attached to a surface and subject to a pulling force on the other end.}
\end{figure}

To obtain exact results for small chains we 
model them as interacting self-avoiding walks (ISAWs) on the square 
lattice \cite{vander} as shown in Fig.~\ref{fig:model}.
Interactions are introduced between non-bonded nearest neighbor monomers.
In our model one end of the polymer is attached to an  impenetrable neutral 
surface (there are no interactions with this surface) while the polymer 
is being pulled from the other end with a force acting along the 
$x$-axis to model the experimental setup.
We introduce Boltzmann weights
$\omega=\exp(-\epsilon/\BC T$) and $u=\exp(-F/\BC T)$ conjugate
to the nearest neighbor interactions and force, respectively, where
$\epsilon$ is the interaction energy,
$\BC$ is Boltzmann's constant, $T$ the temperature and $F$ the
applied force. In the rest of this study we set $\epsilon=-1$ and $\BC=1$.
We study the finite-length partition functions
\begin{equation}
Z_N(F,T)  = \!\!\!\!\! \sum_{\rm all \ walks} \!\!\!\!\!\!\! \omega^m u^x 
     \! =  \sum_{m,x} \! C(N,m,x)  \omega^m u^x,
\end{equation}
where $C(N,m,x)$ is the number of ISAWs of length $N$ having $m$ nearest 
neighbor contacts and whose end-points are a distance $x=x_N-x_0$ apart.
The partition functions of the {\bf CFE}, $Z_N(F,T)$, and 
{\bf CDE}, $Z_N(x,T)= \sum_{m} C(N,m,x) \omega^m$, are 
related by $Z_N(F,T)  =  \sum_{x} Z_N(x,T) u^x$.
The free energies are evaluated from the partition functions 
\begin{eqnarray}
G(x)= -T \log Z_N(x) \; \; {\rm and } \; \;
G(F)= -T \log Z_N(F).
\end{eqnarray}
Here $\langle x \rangle = \frac{\partial G(F)}{\partial F}$ and 
$\langle F \rangle = \frac{\partial G(x)} {\partial x}$ are the 
control parameters of the {\bf CFE} and {\bf CDE},
respectively.

\begin{figure}[t]
\includegraphics[width=3.2in]{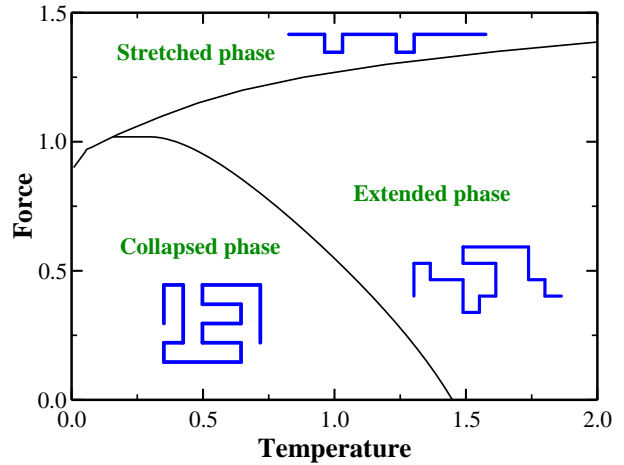}
\caption{\label{fig:phase} The phase diagram  for flexible chains.}
\end{figure}

We enumerate all possible conformations of the ISAW by exact enumeration 
techniques. The major advantage of this approach is that the complete
finite-length partition functions can be analyzed exactly. 
Furthermore scaling corrections can be taken into account by suitable extrapolation 
schemes enabling us to obtain accurate estimates in the thermodynamic
(infinite length) limit \cite{gut}. To achieve a similar degree of accuracy using 
Monte Carlo simulations one typically has to use chains at least two orders 
of  magnitude longer than in the exact enumerations \cite{singh}.  
The greatest challenge facing exact enumerations is to increase the chain length. 
Until now most  exact results for models of small proteins were confined 
to chain lengths of 30 or so \cite{maren,kumar}. 
Here the number of  ISAWs was calculated using
transfer matrix techniques \cite{jensen}. Combined with parallel processing, we
were able to almost double the chain length. To be precise we calculate 
the partition functions up to chain length 55.

In Fig.~\ref{fig:phase}, we show the force-temperature phase diagram
for flexible chains. The qualitative phase diagram remains largely the
same \cite{kumar}. The observed re-entrant behaviour can be explained 
by a ground state with non-zero entropy \cite{kumar,ret}. 
However, the extension up to length 55, allowed us
to obtain a new transition line from the extended state to the stretched
state which is solely induced by the applied force. In contrast to the
lower phase boundary (collapse transition), where the force decreases 
with temperature, the upper phase boundary (stretching transition) 
shows that the force increases with the temperature.

\begin{figure}[t]
\includegraphics[width=3.3in]{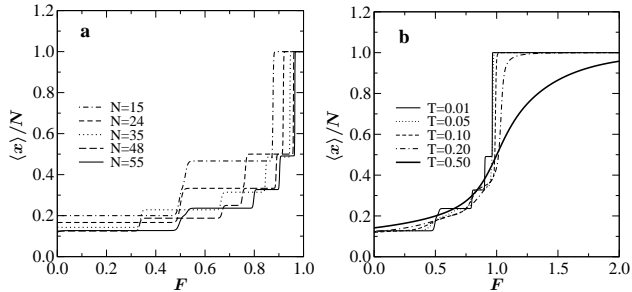}
\caption{ \label{fig:AveX}
The average scaled elongation $\langle x \rangle/N$ vs $F$ at $T=0.01$ 
for various lengths (a) and temperatures at length $N=55$ (b).}
\end{figure}

First we study the model in the {\bf CFE}.
In Fig.~\ref{fig:AveX}a, we plot the average scaled elongation with force 
for different chain lengths at low temperature. Experimentally several
transitions were found  in the force-extension curves corresponding to
many intermediate states \cite{rief2,haupt,hanbin,lemak}. This phenomenon is clearly
confirmed by our study. It has been argued that in the limit of infinite 
chain length, the intermediate states should  vanish  and there will be 
an abrupt transition between a folded and fully extended state \cite{maren}. 
Evidence to this effect can also be seen in Fig.~\ref{fig:AveX}a where we note
that the plateaus at an extension around $0.2$ tends to increase with $N$
while the other plateaus tend to shrink with $N$ (this is particularly so
for the plateau around $0.5$ corresponding to a simple zig-zag pattern of the chain).
Note however, that as we change the chain length from $25$ to $55$, we find 
more and more of these intermediate states. This has also been observed in recent
experiments \cite{rief2,haupt} where the globule deforms into an ellipse and then into a
cylinder. At a critical extension the polymer undergoes a sharp
first order transition into a ``ball string" conformation \cite{rief2,haupt}. This 
shows that finite size effects are crucial in all the single 
molecule experiments and can be seen even for long chains \cite{lemak}.
A simple theoretical argument for the observed behavior is that at low temperature,
where the entropy $S$ (per monomer) of the chain is quite low, the dominant 
contribution to the free energy
\begin{equation}
G =  N \epsilon - \sigma(N,F) \epsilon - N T S
\end {equation}
is the non-bonded nearest-neighbor interaction $N\epsilon$.
The second term is a surface correction and it vanishes in the
thermodynamic limit. However for finite $N$, the system has many
degenerate states depending upon the shape of the
globule. This leads to a surface correction term  $\sigma(N,F)$
which is a function of $N$ and $F$. If $F=0$ the
shape of the globule is like a square and the surface
correction term $\sigma(N,0)$ will be minimized and
equal to $2 \sqrt{N}$. In the {\bf CFE}, there is a
force induced additional contribution proportional to
the extension of the globule which along with $\sigma(N,F)$
stabilizes the intermediate states. When the temperature increases the 
multi-step character of the force-extension curve is  washed out due to
increased contributions from entropy. This effect can be clearly seen in
Fig.~\ref{fig:AveX}b  where we have plotted force-extension curves at 
different $T$.

\begin{figure}[t]
\includegraphics[width=3.3in]{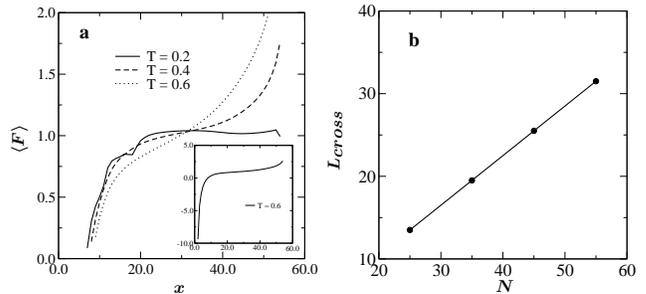}
\caption{ \label{fig:AveF} 
Plot of the average force $\langle F \rangle$ vs the elongation $x$ at 
various temperatures $T$ for $N =55$ (a) and the cross-over length vs $N$ (b).}
\end{figure}

Next we study the model in the {\bf CDE}.
The force-extension curve shown in the insert of 
Fig.~\ref{fig:AveF}a has interesting 
features. It shows that when the distance between the first
and the last monomer (where force is applied) is less than 
the average size of the coil (without force), one needs a
compressing force instead of a pulling force.
The qualitative behavior is similar to one observed in experiments \cite{sevick}
and computer simulations \cite{jorge}.
Since most models do not include confinement in their description, 
such behavior could not be predicted.  
In Fig.~\ref{fig:AveF}a, we show the response of the force when the elongation exceeds
the average size of the polymer. The flat portion of the 
curve gives the average force needed to unfold the chain.
Such plateaus have been 
seen in experiments \cite{haupt,lemak,hanbin}. From  Fig.~\ref{fig:AveF}a one 
can also see  that
the force required to obtain a given extension initially decreases with temperature. 
But beyond a certain extension (close to 30 in this case) the required force
increases with temperature. We note that the curves 
cross each other at a `critical' extension for any temperature (below
the $\theta$-point). We identify this as a cross-over point. In 
Fig.~\ref{fig:AveF}b we plot the position of the cross-over point 
$L_{cross}$ as a function of the length $N$ of the polymer chain. 
We see that the crossover extension increases linearly with the chain length.
This shows that above this point the chain acquires the conformation of the 
stretched state.
The increase in force with temperature generates a tension 
in the chain sufficient to overcome the entropic effect. Since the contribution to 
the free energy from this term is $TS$ ($S$ being the entropy), more force 
is needed at higher $T$ as seen in experiments. Our exact analysis for 
finite chain length shows that applying a force at first favors 
taking the polymer from the folded state to the unfolded state. However, 
rupture or second unfolding occurs when the tethered or unfolded chain attains the 
stretched state, and one requires more force at higher temperature.

\begin{figure}
\includegraphics[width=3.2in]{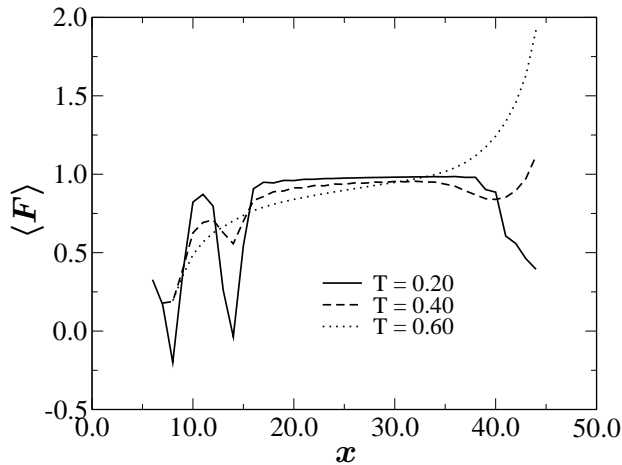}
\caption{\label{fig:AveF-semi} Plot of $\langle F \rangle$ vs $x$ 
for a semi-flexible chain with bending energy $\epsilon_b=0.3$
at different $T$ for $N =45$}
\end{figure}

We model semi-flexible polymers by associating a positive energy 
$\epsilon_b$ with each turn or bend of the walk \cite{kumar}. 
The corresponding Boltzmann 
weight is $\omega_b=\exp(-b\epsilon_b)$, where $b$ is the number of bends
in the ISAW. We again enumerate all walks, but because of
the additional parameter $\omega_b$, we were restricted to $45$ steps.
For a semi-flexible polymer chain, a stretched state may be favored by 
increasing the stiffness. The phase diagram for semi-flexible 
chains is now well established. It has three states namely
(i) an open coil state at high temperature, (ii) a molten globule
at low temperature and low stiffness and (iii) a 'frozen' or 'folded'
state at low temperature and large stiffness \cite{grass,garel,kumar}. 
We note that while the flexible and semi-flexible $F-T$ phase-diagrams 
are qualitatively similar \cite{kumar}, the re-entrant behavior is suppressed
because of stiffness and becomes less pronounced with
increasing bending energy.
In the {\bf CFE}, the probability distribution of the end-to-end
distance has "saw-tooth" like behavior corresponding to
intermediate states during unfolding \cite{kumar}. Therefore, it is important
to study the effect of stiffness on force-extension curves in the 
{\bf CDE}. 
The force extension curves shown in Fig.~\ref{fig:AveF-semi} have striking 
differences to the flexible ones. At low temperatures we see
strong oscillations which vanish as the temperature is increased. Since the
polymer chain has ``frozen conformations" like $\beta-$sheets (the zero-force limit 
of which describes zig-zag configurations inscribed in a square) \footnote{In 3D
[V. A. Ivanov {\em et al}, Macromol. Th. Sim. {\bf 9}, 488 (2000)] a toroidal 
state dominates for stiff chains 
at low temperatures with no applied force. In our (2D) model toroidal states are 
topologically impossible. However the $\beta-$sheet conformations are analogous to
toroidal states.}, it
takes more force to unfold a layer. When about half a layer
has been opened, the bending energy favors a complete stretching of
the layers and hence the force decreases. This phenomenon
allows us to probe a molecule like Titin which has similar 
$\beta$-sheet structure \cite{rief1}.

To summarize, we have presented the exact solution of a model of long (finite)
polymer chains, of direct relevance to recent experiments on the elastic
properties of single biomolecules. The model takes into account several
constraints imposed by the experimental setups: geometric constraints,
excluded volume effects,
attraction between chain segments, finite but large chain length (here up
to $N=55$). It permits one to choose the thermodynamic ensemble ({\bf CFE} 
or {\bf CDE}) dictated by the experimental protocol.
The exact enumeration data permits us to access all parameter values, including
biologically relevant low temperatures where previous studies have failed.
Our results correctly reproduce several experimentally observed effects:
multiple transitions during unfolding, ``saw-tooth'' like oscillations in the
force-extension curve of semi-flexible chains and first-order transition into
a ``ball string'' conformation. Finally, we have identified cross-over
behavior that provides a unified treatment of both pulling and compressing forces 
in the {\bf CDE}.

We would like to thank MPIPKS, Dresden, Germany (SK),
the Australian 
Research Council (IJ,AJG), and the Indo-French Centre for the Promotion 
of Advanced Research (CEFIPRA) (JLJ).
We used  the computational resources of APAC and VPAC.

\end{document}